\documentclass{svproc}
\usepackage{url}

\usepackage{graphicx}
\usepackage{tikz}
\usepackage{comment}
\usepackage{amsmath,amssymb} 
\usepackage{color}

\usepackage[accsupp]{axessibility}  

\usepackage{times}

\usepackage{soul}
\usepackage{url}
\usepackage[hidelinks]{hyperref}
\usepackage[utf8]{inputenc}
\usepackage[small]{caption}
\usepackage{graphicx}
\usepackage{amsmath}
\usepackage{booktabs}
\usepackage{subfigure}

\begin{document}
\mainmatter              
\title{GAN-based generative modelling for dermatological applications -- comparative study}
\titlerunning{GAN-based generative modelling for dermatological applications}  
%
\author{Sandra Carrasco Limeros\inst{1,2} \and Sylwia Majchrowska\inst{1,2} \and Mohamad Khir Zoubi\inst{1} \and Anna Rosén\inst{1} \and Juulia Suvilehto\inst{1} \and Lisa Sjöblom\inst{1} \and Magnus Kjellberg\inst{1}}
\authorrunning{S. Carrasco Limeros, S. Majchrowska et al.} 
%
\tocauthor{Sandra Carrasco Limeros, Sylwia Majchrowska, Mohamad Khir Zoubi, Anna Rosén, Juulia Suvilehto, Lisa Sjöblom, Magnus Kjellberg}
\institute{Sahlgrenska University Hospital, Blå stråket 5, 413 45 Göteborg, Sweden,\\
\and
AI Sweden, Lindholmspiren 3-5, 402 78 Göteborg, Sweden \\
\email{\{sandra.carrasco, sylwia.majchrowska\}@ai.se}}

\maketitle              

\begin{abstract} 
The lack of sufficiently large open medical databases is one of the biggest challenges in AI-powered healthcare. 
Synthetic data created using Generative Adversarial Networks (GANs) appears to be a good solution to mitigate the issues with privacy policies. 
The other type of cure is decentralized protocol across multiple medical institutions without exchanging local data samples.
In this paper, we explored unconditional and conditional GANs in centralized and decentralized settings.
The centralized setting imitates studies on large but highly unbalanced skin lesion dataset,
while the decentralized one simulates a more realistic hospital scenario with three institutions.
We evaluated models' performance in terms of fidelity, diversity, speed of training, and predictive ability of classifiers trained on the generated synthetic data.
In addition we provided explainability through exploration of latent space and embeddings projection focused both on global and local explanations.
Calculated distance between real images and their projections in the latent space proved the authenticity and generalization of trained GANs,
which is one of the main concerns in this type of applications.
The open source code for conducted studies is publicly available~\footnote{\url{https://github.com/aidotse/stylegan2-ada-pytorch}}.
\end{abstract}

\section{Introduction}

In recent years, the use of neural networks has become a very popular and attractive topic for many medical researchers. The applications typically lie in diagnostic support, as one of the key promises of using Artificial Intelligence (AI) in healthcare is its potential to improve diagnosis. The research carried out is related to both tabular data in the form of electronic health records~\cite{bib:AYALASOLARES2020103337}, as well as various types of medical images~\cite{bib:10.1007/978-3-030-11723-8_9,bib:MartinSam}. However, to create reliable deep learning (DL) algorithms that are able to identify complex patterns of medical conditions, they must be trained on a large amount of data. In addition, it is desirable for the model to have a diverse range of cases, as data from a single source may be biased by the acquisition protocol or the population~\cite{bib:neuro,bib:ISICCls22}.

Unfortunately, preparation and annotation of medical data is a costly procedure that demands the assistance of medical specialists. Additionally, access to medical data requires a lengthy approval process due to patient privacy concerns. This makes it almost impossible for different institutions to share data and thus expertise with one another. Although there are some high quality open access dataset initiatives~\cite{bib:mimic-iv,bib:ISIC20}, there is still a great need for much more diverse and complex databases to effectively apply DL.

One way healthcare providers have shared data is by de-identifying or anonymizing the records prior to sharing. But it has its limitations: it is often not private, and it has been reported numerous privacy breaches and re-identification being possible on such datasets~\cite{reidentification}. Synthetic data, on the other hand, is data created from scratch and cannot be traced back to any individual if modeled properly.
Artificial data can be used in two ways - firstly as extensions of small and unbalanced datasets (e.g., of rare diseases) and secondly for anonymization purposes (to replace instead of augment real samples). In both scenarios, synthetic medical data must accomplish two competing goals. The data should accurately reflect the real data and simultaneously offer strong privacy protection for the individuals whose records were used to create it.
 
The main contribution of this work is a detailed study of GAN-based artificial data generation in the case of one of the latest and largest open source databases of skin lesions, namely, International Skin Imaging Collaboration (ISIC) 2020~\cite{bib:ISIC20}. Our research is based on StyleGAN2 with adaptive discriminator augmentation (ADA)~\cite{bib:stylegan2-ada} architecture, which is considered the state-of-the-art in image generation.
We conducted an extensive latent space analysis of the generated images to better understand the structure of the real and synthetic images for the subsequent binary classification task (benign and malignant). In that analysis we compared predictions made on central and edge cases, i.e. those images that do not clearly belong to either of the classes. Edge cases are interesting as they tend to exhibit more variation than clear-cut cases and be more difficult for clinicians to classify.

In this case we have access to a large amount of data in order to generate realistic and diverse synthetic samples. But in a more realistic scenario, we would not have access to such amount of data in just a single hospital. Therefore using generative modeling technique which requires large enough dataset can be difficult. Additionally, we would be constrained by the exchange of data between different institutions with the lengthy and arduous legal procedures that this entails. The different hospitals would have a slightly different distribution, due to the difference in patients, instrumentation, lighting conditions, etc. 
In order to tackle this issue, we simulate a scenario with three hospitals with a different amount of data each. We propose to use Federated Learning (FL) with the aim to synthesise a more complex, fair and diverse dataset through the collaboration of the different institutes, without the need of sharing the data among them, only sharing the parameters of the model.

An overview of the current work for deep learning generation of skin lesions is described in Section~\ref{sec:related_works}. Section~\ref{sec:experiments_design} illustrates our experimental setup for the generation of synthetic medical images with an evaluation of their usefulness and basic characteristics of the examined data. More specifically, we described the used data in detail and provided the training details of the chosen neural networks. In Section~\ref{sec:results} we report the obtained results. Finally, works are summed up, conclusions are drawn and future work is outlined in Sections~\ref{sec:discuss} and \ref{sec:conclusions}.

\section{Literature Review}
\label{sec:related_works}
Generative neural models are becoming more widely used in medical sciences, e.g., generation of mammograms for radiology education~\cite{mamgan}, or realistic chest X-ray images~\cite{chestgan}. In dermatology, GANs are mostly used for automated skin cancer classification~\cite{bib:9062473,bib:GAN21,melanomagan}. Usually, the models are employed as a complementary technique to traditional data augmentation methods, since they increase the amount of training samples and balance the existing dataset. On the other hand, artificial images can serve as the only source of training data to train DL neural networks, which makes data sharing between different institutions easier~\cite{bib:ganrew}.

Most recent works in the field focus on the improvement of classification performance of skin lesion images, which can be achieved with high resolution and rich diversity of generated samples. In these cases, the researchers took into account different real:synthetic data ratios, training subset sizes, and even synthetic sampling techniques~\cite{bib:GAN21,bib:9062473,bib:bissoto2018skin}.
Despite so many possibilities to generate and use synthetic data, the authors~\cite{bib:ganrew} noticed that many of the proposed studies exhibit some flaws and reported that the improvement on accuracy can be the result of the chosen evaluation procedure and test data characteristics.

Another path taken by researchers is the simulation of research institutes using decentralised AI (DAI) to overcome privacy concerns. In healthcare contexts, the most popular technique is a swarm learning~\cite{bib:hpe} scenario (also called peer-to-peer), where no central server is involved to exchange information~\cite{bib:fan2021fairness,bib:bdair2021semisupervised}. In the case of a FL setup we have an additional server to exchange the insights using a chosen aggregation method. These types of experiments demonstrate that DAI improves both performance and fairness compared to central training. On the other hand, in the case of image generation~\cite{bib:rajotte2021reducing}, it was shown that clients with limited and biased data could benefit from other sites while keeping data from all sources private.

\section{Methodology}
\label{sec:experiments_design}

\begin{figure}[h]
\centering
\includegraphics[width=\linewidth]{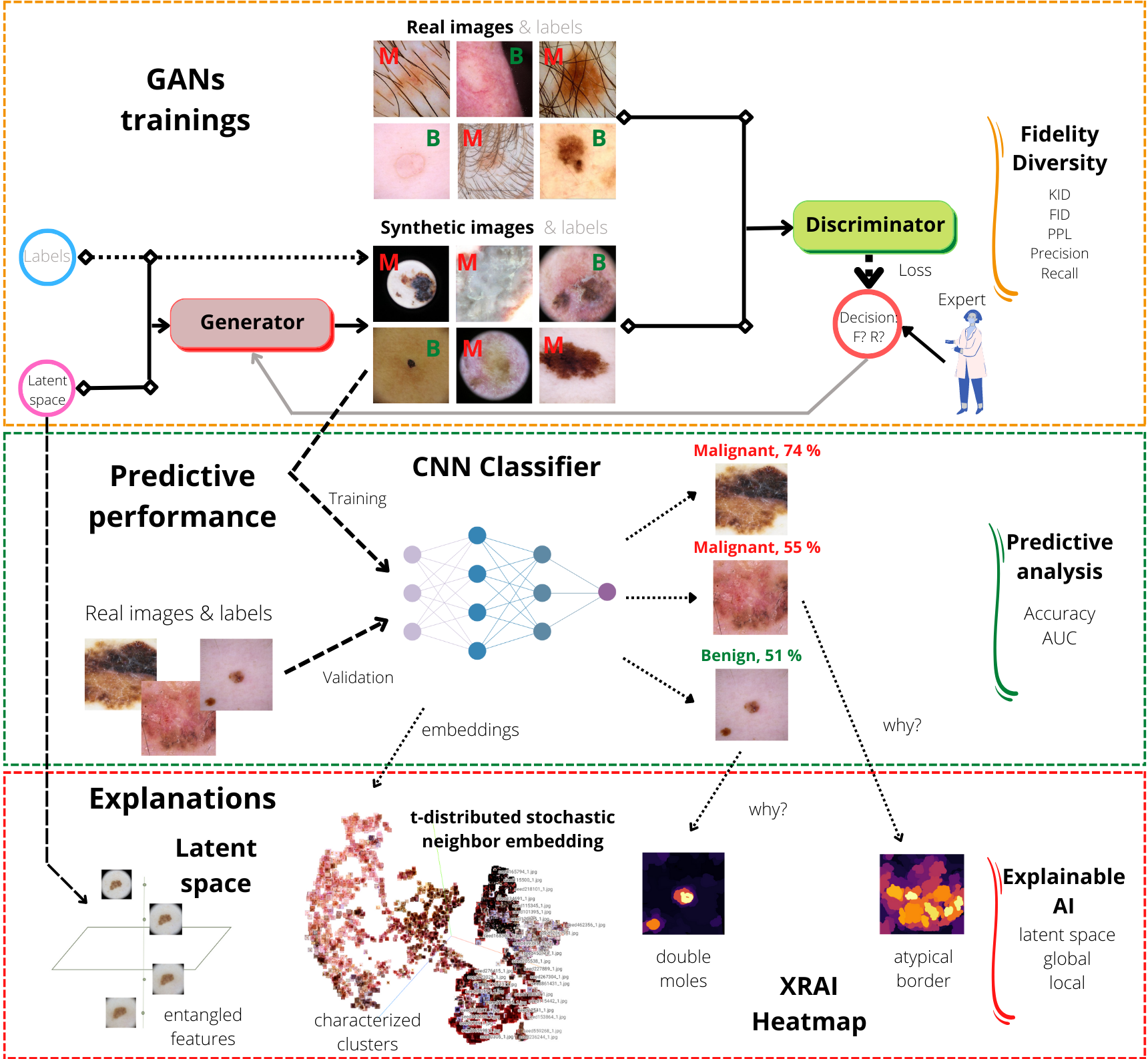}
\caption{A pipeline of a three-stage procedure to generate and validate synthetic data of skin lesions.}
\label{fig:abstract}
\end{figure}

The main aims of our study are to explore and evaluate generative modelling techniques for generating synthetic medical data and to inspect its impact on modelling performance. As personal integrity and regulatory issues are central when using personal health data, we chose to test the generation of artificial data by training models in two different ways: centrally, on a large amount of open access data, and in a hospital scenario (using FL) with 3 independent clients (hospitals), where only model insights are shared between the clients. First, we evaluated the results using the most popular objective metrics for measuring quality and diversity of the generated data, as well as a subjective metric by receiving reliability measurements from experts in the field. Second, we trained classifier on different data configurations for both unconditional and conditional GANs: classical scenario (using only real data), augmentation approach (both real and synthetic data), and GANs-only (only synthetic data). In further procedure we tested trained model performance on the same subset of the real samples and evaluate the similarity in their distributions. Subsequently, we projected into the latent space the real and fake samples to inspect their main characteristics and assess the authenticity of the generative model. Whole pipeline is presented in Fig.~\ref{fig:abstract}.

\subsection{International Skin Imaging Collaboration Database}
\label{subsec:dataset}

In our experiments, the reference dataset for real images is based on the training set of the ISIC 2020 challenge~\cite{bib:ISIC20} extended by malignant cases from previous years' competitions~\cite{bib:kaggle2020}. The whole ISIC dataset collection is a leading standard open access database for medical image analysis using DL techniques. 

Although the ISIC database contains, by the standards of medical data, an impressive amount of datapoints, there are some challenges related to the data quality and curation~\cite{bib:ISICCls22}.
First of all, the dataset is highly unbalanced -- only around 2\% of its samples belong to histopathologically confirmed malignant melanoma cases.
Moreover, the dataset has significant variation in terms of image acquisition set-up,  colors, and sizes. Cassidy et al.~\cite{bib:ISICCls22} have identified a few artifacts which may represent a bias in the dataset, which further can affect the robustness of trained models.
The artifacts can be divided into two categories -- those coming from the patient (e.g., hair, skin tone) and those arising from the acquisition setup (e.g., dermatoscopic frames, clinical markings, stickers, physical rulers, pockets of air, and even removal of part of the lesion).

The considered ISIC database consists of the 37~648 images -- the whole ISIC 2020 dataset, adding 4522 malignant samples from ISIC 2019 -- where 20\% were used for validation in first phase of central trainings.
Later, we splitted the training subset based on patient ID attributes.
To make the FL setup more appropriate, we ensured that the data from an individual patient would not be present on more than one client. For this setup, we created 3 clients and for them, data subsets with 2k, 12k, 20k images respectively. For each client the proportion of malignant and benign was roughly the same as in the whole dataset. In all experiments, we resized the input images to 256 × 256 pixels and normalized them using the same means and standard deviation per channel value as for the ImageNet dataset.

\subsection{Training details}
\label{subsec:trainingdetails}
We investigated StyleGAN2-ADA performance using an original implementation from NVIDIA Research group\footnote{\url{https://github.com/NVlabs/stylegan2-ada-pytorch}}. The ADA mechanism significantly stabilizes training in limited data regimes which we faced in the case of malignant melanoma samples.
We trained StyleGAN2-ADA models with each of the two classes of training set as input, as well as in a conditional setting with and without augmentations.
To select the best model, we considered both the Fréchet Inception Distance (FID)~\cite{bib:FID} and Kernel Inception Distance (KID)~\cite{bib:KID} metrics, along with training speed, similarly as proposed in~\cite{bib:ganrew}.
At the end of this phase, we picked the best GAN to generate the corresponding dermatoscopy images for malignant and benign cases and continued the analysis of the generated synthetic data based on latent space exploration. For the projection of real images into the latent space of the generative model, we needed to optimize the latent code. For this purpose, we used VGG16 model as a feature extractor since this is the one used by default in the StyleGAN2 model by NVIDIA.

The classification task was performed using EfficientNet-B2 model~\cite{bib:efficientnet2019}, pretrained on ImageNet, with Ross Wightman's implementation~\footnote{\url{https://github.com/rwightman/efficientdet-pytorch}}. We chose this model based on the Kaggle competition scoreboard, and the model's well-known performance on skin lesion analysis. As the main goal of the paper was examination of different GANs performance and the classification results on real ISIC2020 are publicly available on the Kaggle platform, the presented results are limited to one classifier. 
However, we employed for centralized scenario also GoogLeNet and ResNet-50 achieving slightly worse results.
During training, we used the Adam method for optimizing the network weights with an adaptive learning rate initialized to 0.0005.
We trained the models for a maximum of 20 epochs and an early stopping with a patience of 3 epochs.
We applied standard data augmentation techniques, such as random rotation, horizontal and vertical flip, during the training phase for all experiments.

For the experiments in a FL setup, we used Flower framework~\cite{bib:beutel2020flower}. We selected Flower framework to build our FL system on due to the simplicity of use, flexibility, ease of adaptation for custom datasets and the possibility to change from a simulated to a truly federated set-up. In our simulated setup, we created a network with 3 clients with different amounts of data and a server, where the weights of the trained model were exchanged every 100 iterations. We used the Federated Average (FedAvg) algorithm~\cite{bib:fedavg} as it is an effective and simple method that is commonly used for federated aggregation, and it is a built-in strategy in Flower.

\subsection{Evaluation procedure}
\label{subsec:metrics}
There are different dimensions that one should take into account when evaluating \break GANs~\cite{Alaa2021HowFI}.
Firstly, fidelity as a measure of reliability, and diversity as a measure of fairness.
Here, FID and KID scores evaluate these two characteristics, but rely on a preexisting classifier (InceptionNet) trained on ImageNet, and  are insensitive to the global structure of the data distribution.
In a similar way, we can look at Precision (P) and Recall (R) scores, where precision measures the fraction of synthetic samples that look realistic (fidelity), while recall measures the fraction of real samples that the generative model can synthesize (diversity). 
In order to evaluate the reliability and realistic look of the artificial images, a survey 
was carried out where experts assessed whether each image surveyed was from a real patient or generated artificially. For this type of subjective evaluation each participant was presented with 200 images roughly split between the classes. All images shown were downsampled to 256 × 256 pixels and generated from the cGAN model.
On the other hand, Perceptual Path Length (PPL)~\cite{bib:ppl} measures whether and how much the latent space is entangled or regularized, in the end, being capable of capturing the consistency of the images.
Secondly, another dimension to look into is predictive performance. Although this is an immediate consequence of the above two concepts, it refers to the fact that samples should be just as useful as real data when used for the same predictive purpose. Here, we built a melanoma classifier to assess the predictive performance when using the synthetic data for training and the real data for testing.

As in the case of medical data, privacy is the most important factor, we evaluated the generalization or authenticity of the generative process~\cite{Alaa2021HowFI}. It measures the rate at which the model invents new samples and is defined as the fraction of generated samples that are closer to the training set than other training data points.
However, despite so many metrics, there is still no clear and gold standard protocol to evaluate the quality of the generated samples.

Finally, since the StyleGAN2-ADA framework learns a linearly separable latent space \cite{bib:stylegan}, it was tempting to explore if image editing is possible by manipulating the latent input of a trained unconditional GAN model on skin cancer images. Different methods were tested with the aim to see if we could obtain directions in the latent space where the influence of one feature could be controlled while keeping the rest of the image intact. Initially, we tried shifting a vector across the separating hyperplane between two features (e.g. dermascopic frame). This method showed qualitatively good results, however, it required using an auxiliary classifier as well as a large sample size of latent vectors. Later, we tried finding the principal directions of a large sample size and visually examined which directions yielded interpretable feature manipulation. Lastly, to mitigate the need to generate a large sample size of latent vectors, we used semantic factorization~\cite{bib:sefa}.
\section{Results}
\label{sec:results}

\subsection{GANs trainings}
In the first phase of our experiments, we established the best model in terms of fidelity and diversity using well-known metrics such as KID, FID, P, R, and PPL (see Table~\ref{tab:gans_metrics}). 
It is worth  noting that the GAN responsible only for malignant melanoma generation (mal-GAN) had around 6 times less data than for benign cases (ben-GAN). Additionally, the vast majority of malignant melanoma examples in ISIC 2020 and ISIC 2019 exhibit a black frame of the dermatoscope, which highly biases the output. We discovered that using global explainability techniques~\cite{kim2018interpretability}, and it was also observed before by other researchers~\cite{bib:mikolajczyk}. We used a conditional setting to provide the model with a wider variety of samples of dermatoscopy images, since there are a lot of characteristics that are common for both classes. We confirmed that this is beneficial for the minority class (malignant), achieving higher FID and KID scores.
For this setting, we used ADA mechanism with and without ($_{w/o\,color}$) color augmentation. The first experiment clearly led to leakage of color augmentation to the generated examples (i.e. a lot of skin samples had an unnatural red or violet hue). In addition, given the amount of data we tested whether in our particular case the augmentation could be detrimental for the training, so we repeated the experiments without any augmentation, obtaining significantly worse results.

We received four responses to the qualitative survey from two dermatologists and two deep learning experts. The overall average accuracy of the participants was 54\%. There was no feature in any image that clearly suggested to the participants that the image was either real or synthetic. All experts were able to correctly label five of the 200 images  as true positives, although they also falsely labeled seven of 200 images as false negatives. The participants expressed that they found it difficult to tell what image belonged to each class. 

\begin{table}[!h]
\centering
\begin{tabular}{lrrrrr}
\hline
Scenario              & KID (\%)& FID  & P    & R     & PPL \\
\hline
ben-GAN               & 0.42   & 7.99  & \textbf{0.77} & \textbf{0.45}  & 60 \\
mal-GAN               & 0.47   & 15.46 & 0.62 & 0.40  & \textbf{51} \\
cGAN                  & 0.32   & 7.33  & 0.75 & 0.42  & 193 \\ 
cGAN$_{w/o\,color}$   & \textbf{0.24}   & \textbf{7.02}  & 0.75 & 0.44  & 101 \\
\hline
\end{tabular}
\caption{Calculated metrics for each of the generative models tested in the centralized setting. The best KID and FID scores were achieved for conditional StylGAN2-ADA without color augmentations (cGAN$_{w/o\,color}$). Unconditional GAN for benign class (ben-GAN) is slightly better in terms of precision and recall. The unconditional models have lower PPL scores, showing a better regularization of the latent space due to the fact that here we modeled only the distribution of one class.}
\label{tab:gans_metrics}
\end{table}

In the current use case, we have access to a large amount of data to generate realistic and diverse synthetic images. However, in a more realistic scenario, one single hospital can have a problem to gather enough amount of samples to work with GANs. To tackle this issue, we simulated a scenario where three hospitals with different amounts of data are connected in a FL setup (see section~\ref{subsec:dataset} for details). 
Using this simulated hospital scenario, we observed faster convergence and improved quality of the generated images mainly for the client with the smallest data resources - it converges around 1.6 times faster achieving a FID score similar to that of the central conditional GAN training.
As the data distributions between different clients only differed in size, we put more emphasis on the classification task with centrally trained models.

\subsection{Predictive performance with classifier}
After the evaluation with general metrics, we performed a study on predictive performance to measure how useful the synthetic data is for the subsequent task, i.e.  malignant melanoma diagnosis (Table~\ref{tab:cls_metrics}). As a baseline for the experiments, we first train the classifier on training subset of the real images of ISIC dataset, and then tested it on the validation set.
Secondly, GAN-based augmentation was performed using two types of GANs models with two scenarios: training on balanced synthetic dataset with 55k images and testing on real (\textit{synt}) validation subset (the same as in baseline experiment) and training on real images adding 22k synthetic melanoma samples (\textit{aug}) to balance the dataset. The introduction of highly underrepresented malignant melanoma cases improves the classification accuracy roughly of few~pp in both scenarios (Table~\ref{tab:cls_metrics}). Overall GAN-based augmentation technique does not provide reliable improvements in case of classification using the whole ISIC 2020 and malignant samples from ISIC 2019.

\begin{table}
\centering
\begin{tabular}{l|rrrrr}
\hline
Scenario                          & Acc           & AUC   \\
                                  & (\%)          & (\%)  \\\midrule
Real-baseline                     & 97.8          & 98,8 \\\midrule
synth-GAN                         & 94.1          & 94.2 \\
aug-GAN                           & 97.8          & 98.6 \\\midrule
synth-cGAN                        & 94.7          & 96.7 \\
aug-cGAN                          & 97.8          & 98.6 \\\midrule
synth-cGAN$_{w/o\,color}$         & 92.6          & 92.7  \\
aug-cGAN$_{w/o\,color}$           & \textbf{97.9} & \textbf{98.8} \\ 
\hline
\end{tabular}
\caption{Calculated metrics for each of the classification scenarios with EfficintNet-B2. 
We generated synthetic samples based on StyleGAN2 using conditional (cGAN) and unconditional (GAN) settings.
Further classifier was trained on clear synthetic (\textit{synth}) balanced 55k dataset, 
or on real images with additional 22k fake malignant lesions images (\textit{aug}) to balance dataset, 
and then in all cases tested on the same real images validation set.
}
\label{tab:cls_metrics}
\end{table}

\begin{figure}[h]
\centering
            \subfigure[] 
            {
                \label{subfig:lab1}
                \includegraphics[width=.55\linewidth]{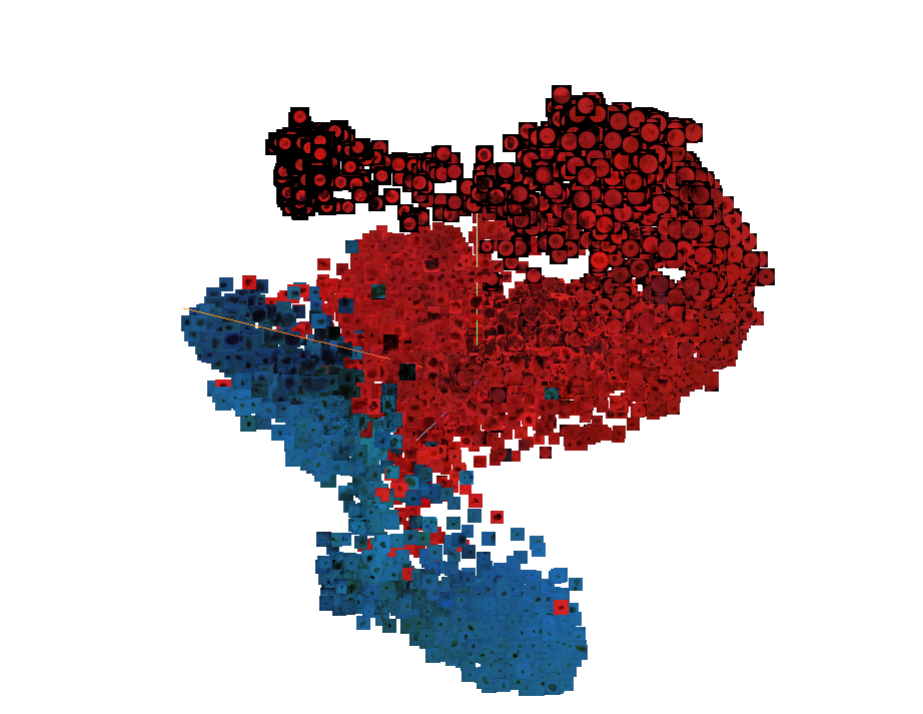}
            } 
            \subfigure[] 
            {
                \label{subfig:lab2}
                \includegraphics[width=.55\linewidth]{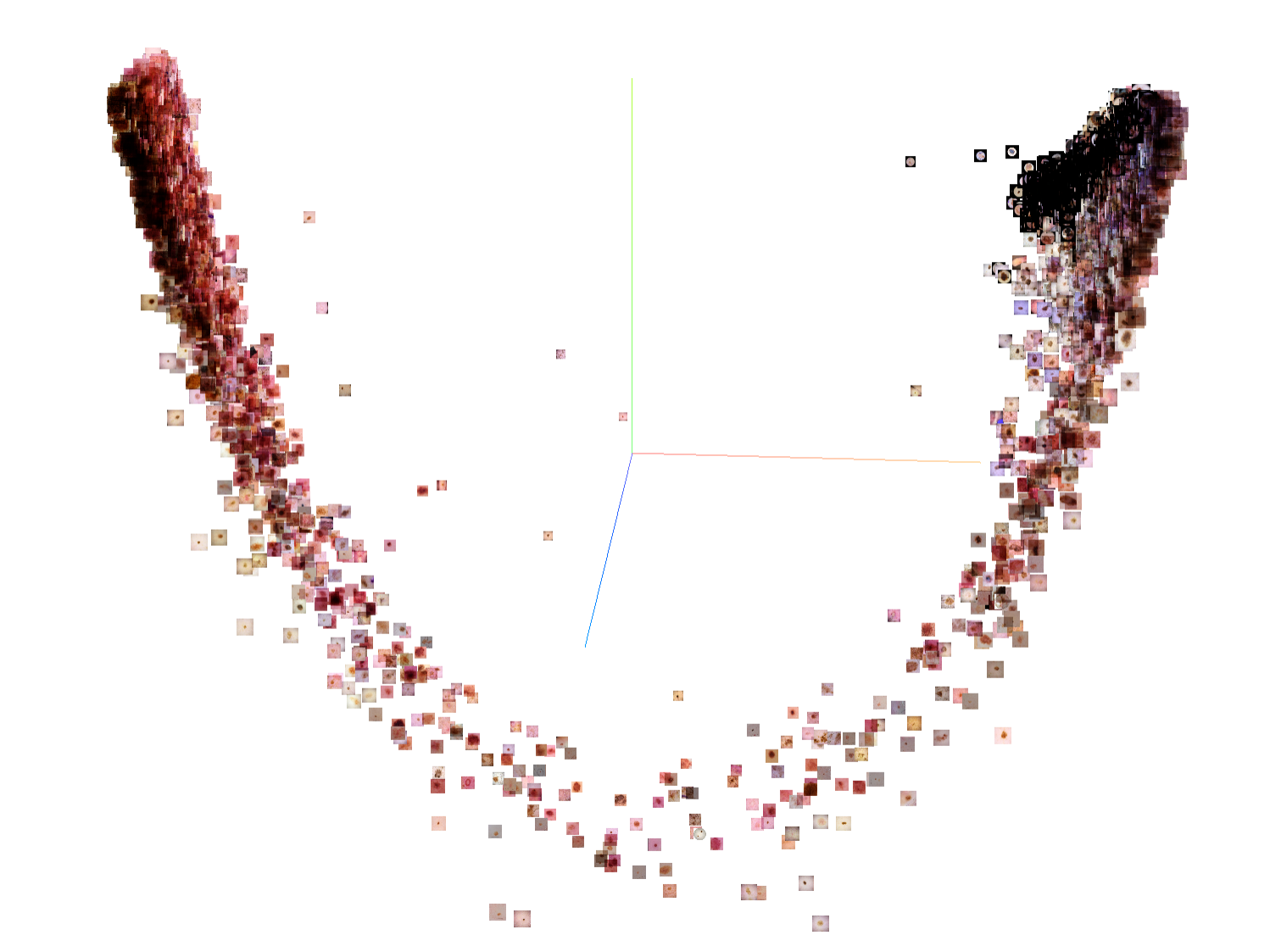}
            }
\caption{Real images projections in the latent space of the generator (a). Projected embeddings of real and synthetic data coming from the classifier trained on synthetic data (b).}
\label{fig:tsne}
\end{figure}

\subsection{Explanations of the predictions}
To measure the authenticity and ensure that our generative model is not copying the real data we performed the following experiment. 
First, we projected 12 thousand samples from the real dataset into the latent space of the generator. This gave us the latent codes that caused our generator to synthesize the most similar output to the input image. To optimize for a latent code for the given input images, we used a VGG16 model as a feature extractor and computed the loss on the difference of the extracted features for both the target image and generated output and performed backpropagation.

Next, we extracted the features of both the real and their projected images using the last convolutional layer of our classifier trained on real and synthetic data. These embeddings were visualized in a 3D space using t-distributed stochastic neighbor embedding (t-SNE) method~\cite{bib:tsne}. This allows visually exploring the closest near neighbors of each real image using cosine distances, and in this way examining how close the real images and their corresponding generated images are. Fig.~\ref{fig:tsne} shows examples of real images projections in the latent space of the generator (with benign marked on red, malignant -- blue) and projected embeddings of real and synthetic data. In both cases there is visible separation between two clusters created by two examined skin lesion classes. However, there are still plenty of the cases in the middle between two clusters, what is visible in Fig.~\ref{fig:tsne}(b), and even mixed with improper class (Fig.~\ref{fig:tsne}(a)). Additionally, we spotted some clusters inside classes, which are associated with instrumental bias, such as ruler and black dermatoscope frame.

\begin{figure}[!h]
\centering
\includegraphics[width=.8\linewidth]{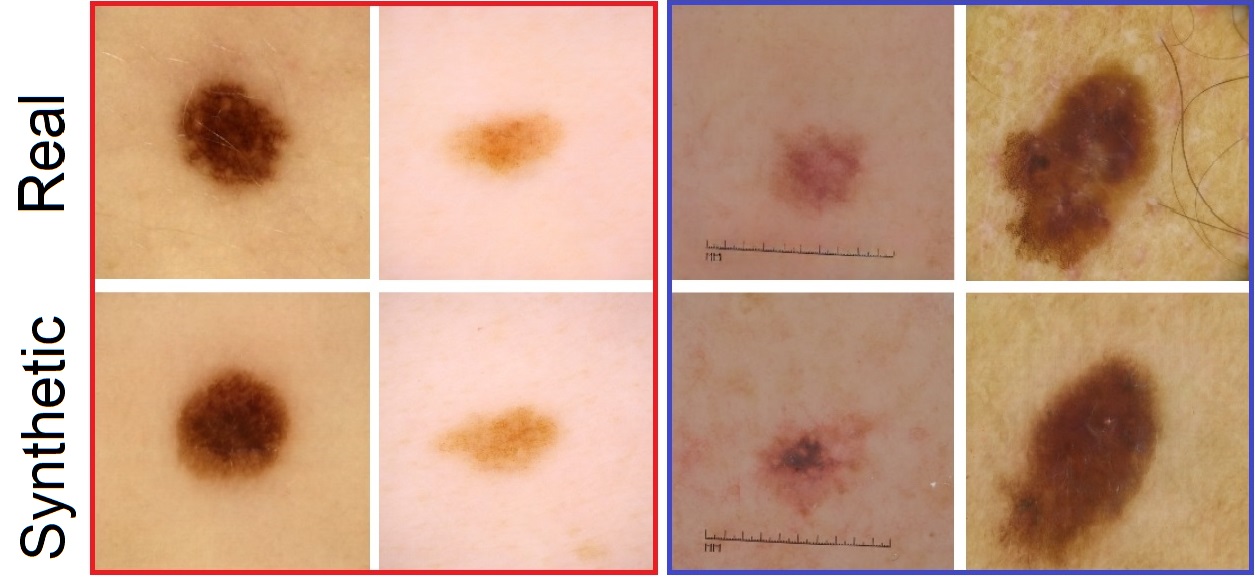}
\caption{A few examples of the closest (red frame) and the most distant (blue frame)  pairs real-synthetic in terms of cosine distance. The closest examples are still not identical, whereas the furthest ones are still similar to the real image we wanted to retrieve.}
\label{fig:distant}
\end{figure}

For a more systematic inspection, we computed the cosine distances between the different pairs of real images and their projected samples. The mean distance was equal to 0.1444 and the median 0.00283 with only two projections being \textit{too close} in terms of \mbox{$Q1=0.013$} (range of 1e-5) to the real images. Only in these two cases the closest neighbor was the projection of the target image, meaning that the generative model could have memorized that sample.
We treated this as a  measure of the authenticity of the generated samples. We also spotted that some of the images were very distant from their projections (around 2) but still resembled the target image (Fig.~\ref{fig:distant}).

\begin{figure}[!h]
\centering
\includegraphics[width=.8\linewidth]{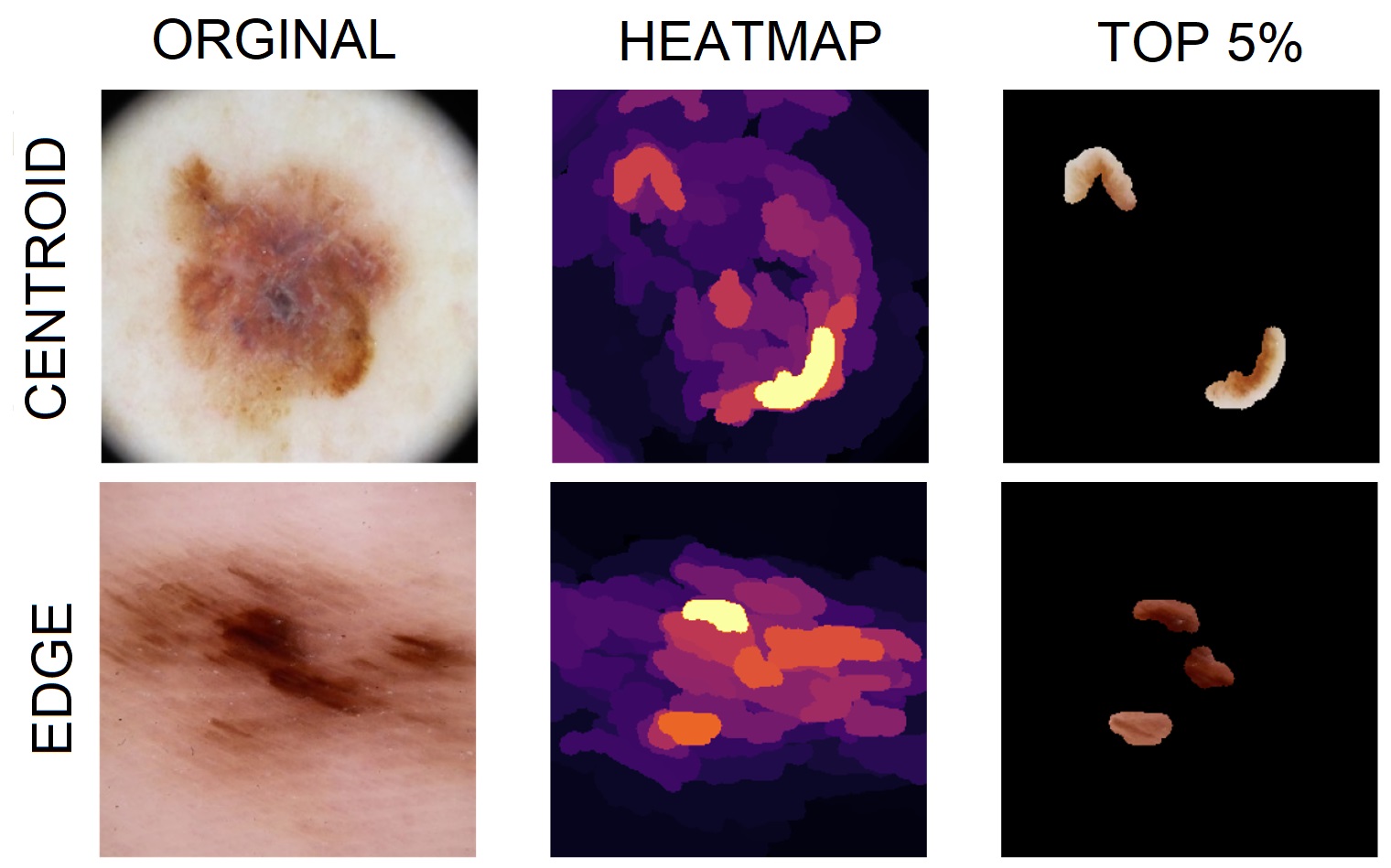}
\caption{Two examples from the malignant class, which were found in the center, and in the boundary between two clusters respectively, examined using XRAI heatmaps.}
\label{fig:xai}
\end{figure}

We also scrutinized the images that lie in the boundary between the two clusters (Fig.~\ref{subfig:lab2}), which correspond to edge cases, i.e. cases that are difficult to classify. These edge cases were studied using local explanations with the XRAI method~\cite{bib:XRAI}, to identify why these samples were more difficult for the network. 
For images of malignant lesions that belong to the centroid of the embeddings, we found that the mole itself is the most important part of the image for the final prediction. In the sample image, the network focuses on boundary pixels which represent asymmetry in the mole, one of the main clues for detecting malignant melanomas. On the other hand, in edge cases the results are not as evident due to image distortions or poorly centred moles (Fig.~\ref{fig:xai}).
We selected all the misclassifications and edge cases and generated N neighbors using a distance of 0.1 to augment the dataset with more complex examples with the aim of making it more robust. First experiments showed an improvement in performance in those edge cases. However, in future a deeper analysis will be performed to confirm this statement. 

\begin{figure}[!h]
    \centering
    \begin{tabular}{ccccc}
         \includegraphics[width=0.15\linewidth]{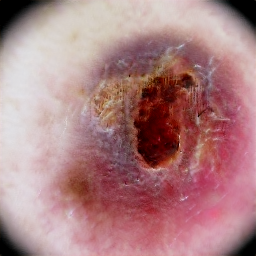} & 
         \includegraphics[width=0.15\linewidth]{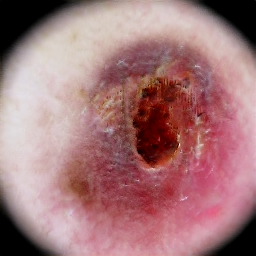} &
         \includegraphics[width=0.15\linewidth]{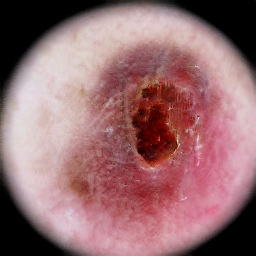} &
         \includegraphics[width=0.15\linewidth]{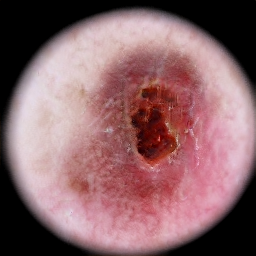} &
         \includegraphics[width=0.15\linewidth]{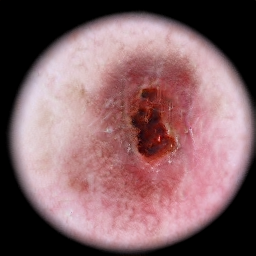}
    \end{tabular}
    \begin{tabular}{ccccc}
         \includegraphics[width=0.15\linewidth]{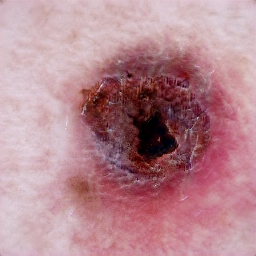} & 
         \includegraphics[width=0.15\linewidth]{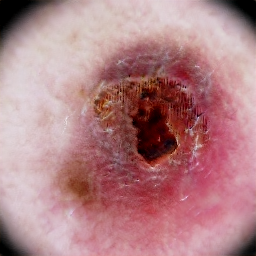} &
         \includegraphics[width=0.15\linewidth]{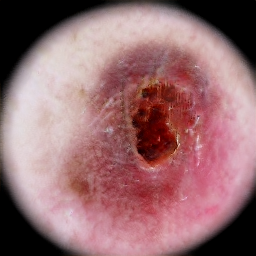} &
         \includegraphics[width=0.15\linewidth]{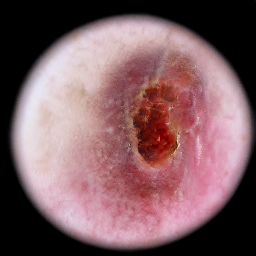} &
         \includegraphics[width=0.15\linewidth]{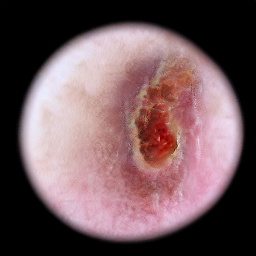}
    \end{tabular}
    \begin{tabular}{ccccc}
         \includegraphics[width=0.15\linewidth]{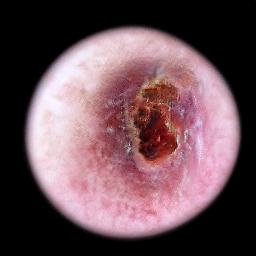} & 
         \includegraphics[width=0.15\linewidth]{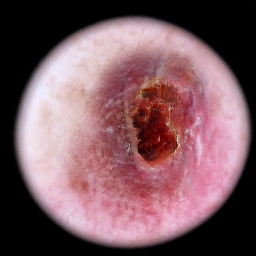} &
         \includegraphics[width=0.15\linewidth]{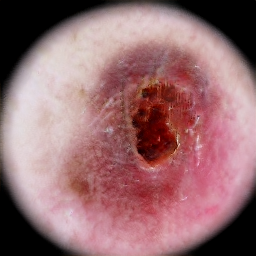} &
         \includegraphics[width=0.15\linewidth]{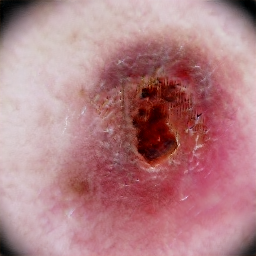} &
         \includegraphics[width=0.15\linewidth]{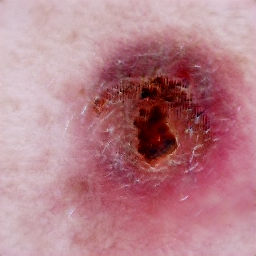}
    \end{tabular}
    \caption{Examples of image editing using the SeFa framework. The image in the middle in all three rows (3rd column) is the original image. The columns correspond to the $\mathbf{w}$-vector being moved and the rows are different directions.} 
    \label{fig:sefa}
    
\end{figure}

Finally, we took care of explanation of trained GANs by providing the manipulations of latent input. 
The features whose influence we focused on reducing were the dermoscopic frames present in malignant melanoma images. We explored the semantic factorization (SeFa) method \cite{bib:sefa}, which does not require a sample size of latent code or auxiliary classifiers. The latent $\mathbf{w}$-vector corresponding to the image in Figure~\ref{fig:sefa} was shifted along the second, forth and sixth eigenvectors, obtained using the SeFa framework.

In Figure-\ref{fig:sefa} we see that applying SeFa image editing suggests less entangled features from visual inspection of the images across different directions. The eigenvectors displayed were chosen from a larger qualitative evaluation of 100 images along the first 10 largest eigenvectors. Visually, the second, fourth, and sixth eigenvectors showed the best result in removing the frames while leaving the other features intact. We observed a better result using this method of image editing compared to the other tested methods (for details, see Section~\ref{subsec:metrics}). 

To assess the quality of the edited images, we first generated a large sample size of images all containing frames. After acquiring the images we removed the frames by shifting the latent vectors along the direction where the presence of the dermoscopic frame was minimized. Finally, we trained a classifier on these images for the malignant melanoma with a training set of 10k images per class and a test set consisting of real images. The resulting test accuracy was $87\%$. 

\newpage
\section{Discussion}
\label{sec:discuss}

In our study, we explored the state of the art DL-based techniques to generate, classify, and explain computed results for skin lesion diagnosis. Our experiments are based on ISIC 2020 and ISIC 2019 datasets, which are one of the largest but very unbalanced open access database. 

Samples generated using different types of GANs and settings exhibits slightly different appearance, which is reflected by different values of calculated evaluating GANs metrics. The PPL measure, which is capable of capturing the consistency of the images, is the lowest for generated malignant melanoma samples by unconditional GAN. However, this is not connected with the lowest KID and FID scores indicating the dissimilarity between two probability distributions (real and fake) using samples drawn independently from each distribution. Lower PPL score is related to the smallest amount of malignant data, and in result more regularized and narrow distribution of latent space. 
The second observation may be connected with the fact, that KID and FID rely on a pre-existing classifier (InceptionNet) trained on ImageNet consisting of different images then skin samples. 
The results also indicates that the cGAN model is prone to generating more realistic looking melanoma (using some features from benign samples) than the mal-GAN. No statistical conclusion can be drawn from the small sample size in the survey where cGAN generated images were used. However, the results do suggest that subjectively, experts are unable to tell an artificial lesion from that of a real patient. There was no specific feature that the experts picked up on in the generated data as an artifact of the model. Therefore, qualitatively the synthetic data pass for real in the eyes of experts.

In case of classification, we have not observed a large improvement of the performance of the classification network based on synthetic data generated by StyleGAN2-ADA. Actually the results archived in different scenarios do not differ much. This may be affected by a large size of the real dataset, but also with fact that some features connected with for example method of collecting data (existence of black dermoscopic frame) may be entangled with specific class.

Performed exploration of the latent space showed that there is a clear separation between the projections of the real and generated samples. Measured distance between the projections of real and the closest synthetic image proved the authenticity of the generated samples. Our main interest in the explanation of classification results focused on the edge cases, as the dermatologists are paying special attention to those cases that lie in the boundary and are not so obvious. We spotted that the network output is often biased by acquisition protocols, as well as some patient-related features. The main issue seems to be the area covered by the mole on the image. However, this topic requires closer examination. Editing images using latent directions could be a useful tool in removing unwanted artifacts from images. However, the dermoscopic frames were only present in malignant melanoma images, so the characteristics of the class labels were entangled with the dermoscopic frames. This entanglement resulted in changes in separate features when removing the frame artifact and did not leave the malignant melanoma data intact. For future steps, using this technique may show promising results in data normalization and generalization in different domains.

On the other hand, as GANs training requires a large investment in computing and data resources, the FL setup may be a solution for smaller institutions with lack of access to sufficient data resources. Achieved results confirmed that the generation of skin lesions in a distributed setup can lead to similar performance with respect to the quality and diversity of generated samples, with a significant faster convergence. However, to make a final verdict on this matter, it is necessary to conduct further research into different aggregation algorithms, privacy preserving techniques, and even defense mechanisms against adversarial attacks.

\section{Conclusions}
\label{sec:conclusions}

GAN-based augmentation is an extensively explored technique for medical imaging applications, especially in the case of very rare diseases. First of all, it helps in the creation of larger and more balanced datasets.
Secondly, it creates non-real data, which can be more easily shared amongst the medical community. However, the results achieved with the addition of synthetic data reported in literature show an improvement in accuracy of only a few percents without clearly explaining the reason. On the other hand, GAN-based anonymization suffers from an unset gold standard in measuring its performance.

To utilize GANs in generating synthetic healthcare data, a number of considerations need to be made. First, one should consider the architecture. In our case, we chose between central unconditional GANs per each class, conditional GAN and FL setup. The usefulness of chosen architectures mainly depend on computational resources and time - unconditional GAN can be good option with small amount of classes due to long duration of training of single GAN. If a massive, annotated dataset exists, training the GAN centrally is preferable but in case of a more realistic scenario of data being siloed in an institution, the benefit from FL is noticeable particularly for smaller institutions. 

Second, the created synthetic data should be inspected from multiple different points of view. Common features to emphasise are fidelity and diversity, which are important to understand how well the synthetic data represents the underlying real data. Importantly, as the goal in healthcare is to avoid sharing data, it is also crucial to inspect the authenticity of the synthetic examples to make sure they are not simply copying the training data. Additionally, the synthetic data should be as useful as the real data for the subsequent task (e.g. classification) and not allow inferences based on features that are not related to the case, but, for example, to the way the data were collected (e.g., linking a black dermatoscope to malignant melanoma).

\section{Acknowledgements}

GANs trainings were conducted during first rotations of the \textit{Eye for AI Program} thanks to the support of Sahlgrenska University Hospital and AI Sweden. The  authors  wish  to express their thanks for other partners: Zenseact, AstraZeneca, and Chalmers University of Technology.

%
%

\end{document}